# Python (deep learning and machine learning) for EEG signal processing on the example of recognizing the disease of alcoholism

Rakhmatulin Ildar, PhD
South Ural State University, Department of Power Plants Networks and Systems
76, Lenin prospekt, Chelyabinsk, Russia, 454080
ildar.o2010@yandex.ru
**https://github.com/Ildaron/3.eeg_recognation**

**Abstract**
Alcoholism is one of the most common diseases in the world. This type of substance abuse leads to mental and physical dependence on ethanol-containing drinks. Alcoholism is accompanied by progressive degradation of the personality and damage to the internal organs. Today still not exists a quick diagnosis method to detect this disease. This article presents the method for the quick and anonymous alcoholism diagnosis by neural networks. For this method, don't need any private information about the subject. For the implementation, we considered various algorithms of machine learning and deep neural networks. In detail analyzed the correlation of the signals from electrodes by neural networks. The wavelet transforms and the fast Fourier transform was considered. The manuscript demonstrates that the deep neural network which operates only with a dataset of EEG correlation signals can anonymously classify the alcoholic and control groups with high accuracy.
On the one hand, this method will allow subjects to be tested for alcoholism without any personal data, which will not cause inconvenience or shame in the subject, and on the other hand, the subject will not be able to deceive specialists who diagnose the subject for the presence of the disease.

## 1. Introduction

According to the World Health Organization, in recent decades the number of patients with alcoholism grew. Research shows that alcohol abuse is associated with behavioral disinhibition, but the neurophysiological mechanisms governing these relationships remain largely unknown. For these reasons, the diagnosis of this disease is difficult. The disease can be identified by the many symptoms. Anuragi et al. [1] and Onarom et al. [2] described the biological processes that occur in the brain during drinking. Ishiguro et al. [3] and Kumar [4] described the physiological consequences of the long-term intake of drinks containing alcohol. These articles showed the complexity of the process operation of the brain during illness and the complexity of diagnosing the presence of this disease. For accurate diagnosis, this disease for specialists needs many private information about patients. But not all patients want to be diagnosed openly. Therefore, the purpose of the research to develop an anonymous method for classifies the alcoholic and control groups by neural networks with the EEG signals dataset.
Today, medicine has stepped far enough in this direction. Winterer et al. [5], Patidar et al. [6] and Acharya et al. [7] provided an overview of the EEG signals of patients diagnosed with alcoholism.

There is enough information in these works to understand the situation in the field of detection of alcoholism by the EEG signal. But interestingly, despite the seeming knowledge of this issue in the EEG field, many papers on this research have conflicting results. Jeremy et al. [8], showed that compared with men, women are at increased risk of negative physical and neurocognitive correlates of alcohol consumption. In research proved that alcohol abuse has a detrimental effect on the dynamics of EEG suppression of the reaction in the theta range. The opposite conclusion is in the work by Ahmadi et al. [11]. Ahmadi decomposed the EEG signal is into five frequency subbands using the wavelet transform. He showed that there is a lower synchronization in the sub-band of beta frequencies and a loss of lateralization in the sub-band of alpha frequencies in alcoholic subjects. Ahmadi realized classification by machine learning algorithms. But in the research, deep neural networks were not used. But, Paulchamy et al. [15] used all threshold alpha, beta, and theta waves to detect this disease in subject.
Ziya et al. [9], implemented software in the Matlab program for classifying by EEG indicator - alcoholic or control subject was presented. The research did not present the result of signal pre-processing. The article does not have enough information about the neural network model. Wajid et al. [10] used EEG data to extract EEG characteristics such as absolute power (AP) and relative power (RP). The classification accuracy of the model is not high. Guohun et al. [12], showed that the areas with electrodes - C1, C3, and FC5 for alcoholic’s groups are significantly different. Mingyue et al. [13] present a new algorithm for analyzing an EEG signal. Mingyue calculated the distinguish non-linear EEG characteristics with alcoholics and controls by the exponential strength Ratio Index (EPRI). But in the research deep neural networks were not used. Joel et al. [15], for the diagnosis of alcoholism by EEG extracted features from four-minute records of EEG of the scalp with eyes closed. In finally the influence of age and gender on the diagnosis of alcoholism was researched. Madhavi et al. [16] noted that increased absolute theta strength in patients with alcohol dependence in all areas of the scalp. Also, Madhavi considered the increase in theta login power in male alcoholics in the central and parietal regions. Anuragi et al. [17] showed that chronic alcoholism is associated with a high frequency of low-voltage recordings.

## 2. Materials and method

In this manuscript, the dataset from Henri Begliter (Laboratory of Neurodynamics at the Center for Health at New York State University in Brooklyn, presented publicly, https://archive.ics.uci.edu/ml/datasets/eeg+database) was used. This dataset from the research of genetic predisposition to alcoholism. In experiment 64 electrodes placed on the scalp were used (frequency of reading signal of 256 Hz).
Two groups of subjects: an alcoholic and a control group were involved in the experiment. Each subject to either one stimulus (S1) or two stimuli (S1 and S2) was subjected. S1 and S2 are a set of images of objects selected from the set of images of 1980 Snodgrass and Vanderwart.
The dataset has the following structure, 480 tables in format - csv for training and 480 tables for verification. The table shows 64 electrodes. Each electrode has 256 records with a duration of 1 second, table. 1.

Table. 1. The structure of the data file for training - Data1, test

| | trial number | sensor position | sample num | sensor value | subject identifier | matching condition | channel | name | time |
|---|---|---|---|---|---|---|---|---|---|
| 5 | 30 | FP1 | 0 | -3.55 | a | S1 obj | 0 | co2a0000364 | 0 |
| 6 | 30 | FP1 | 1 | -5.015 | a | S1 obj | 0 | co2a0000364 | 0.003906 |
| 7 | 30 | FP1 | 2 | -5.503 | a | S1 obj | 0 | co2a0000364 | 0.007813 |

The table contains the following information: sensor position, sensor value (µV), subject identifier (Alcoholic(a) or Control (c)), matching condition, name(a serial code assigned to each subject), time(inverse of sample num measured in seconds))

The location of the sensors on the head is shown in fig. 1.

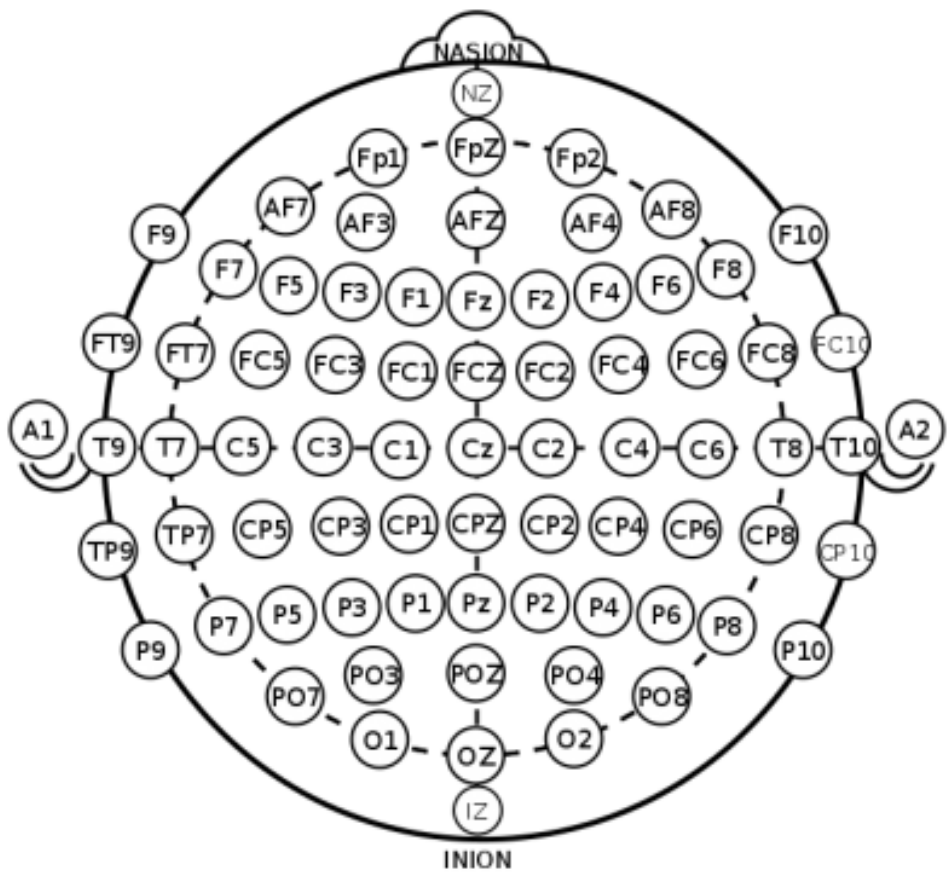

Fig. 1. Arrangement of 64 electrodes during EEG recording

Fig. 2 demonstrates the dataset with the library plotly, python3.7, where a - alcoholic group with picture from collection - s1 (alk_s1), b - alcoholic group with picture from collection - s2 (alk_s2), c - not alcoholic group with picture from collection-s1 (not_alk_s1), d - not alcoholic group with picture from collection-s2 (not_alk_s2).

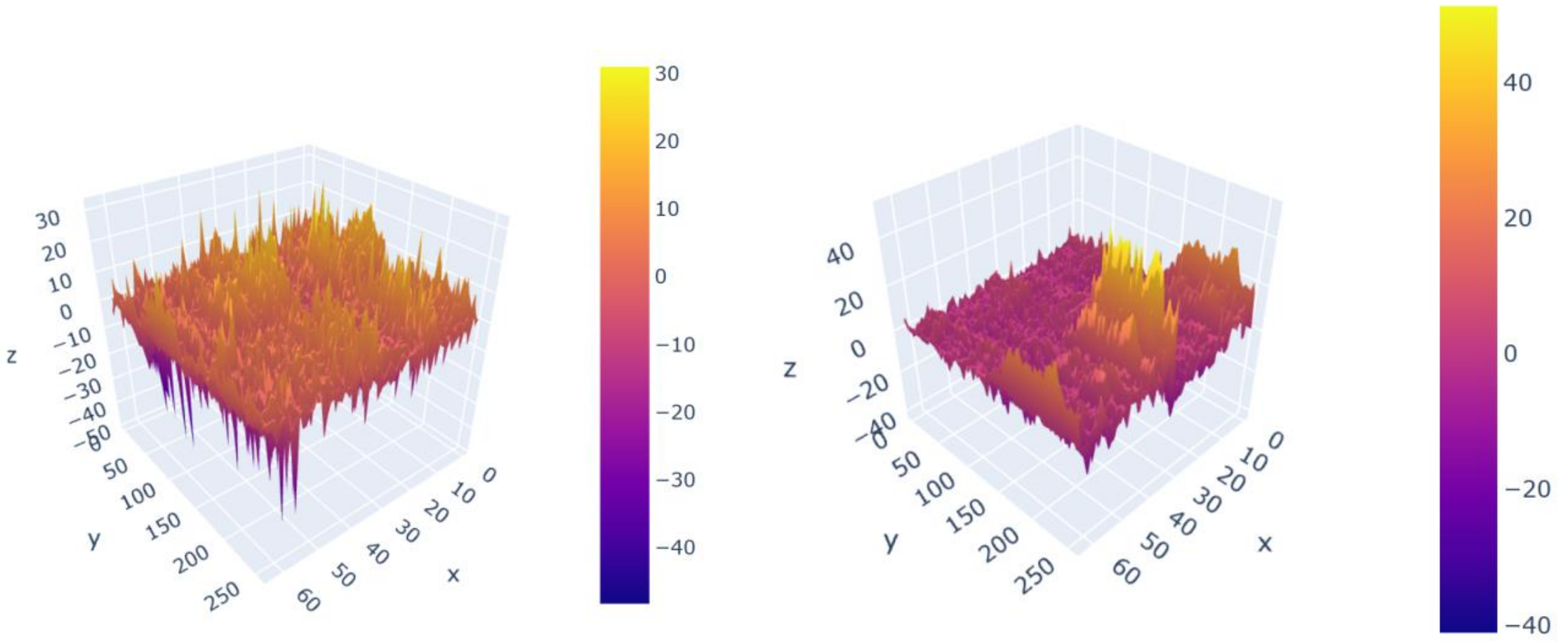

a b

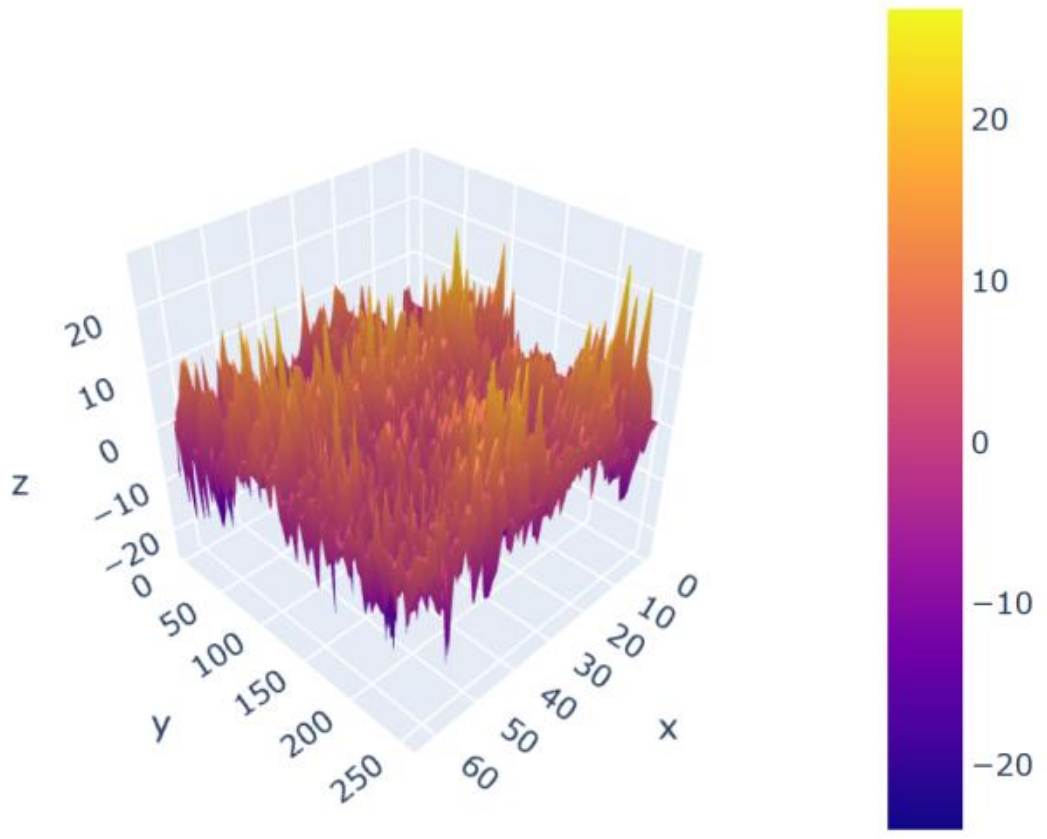

c

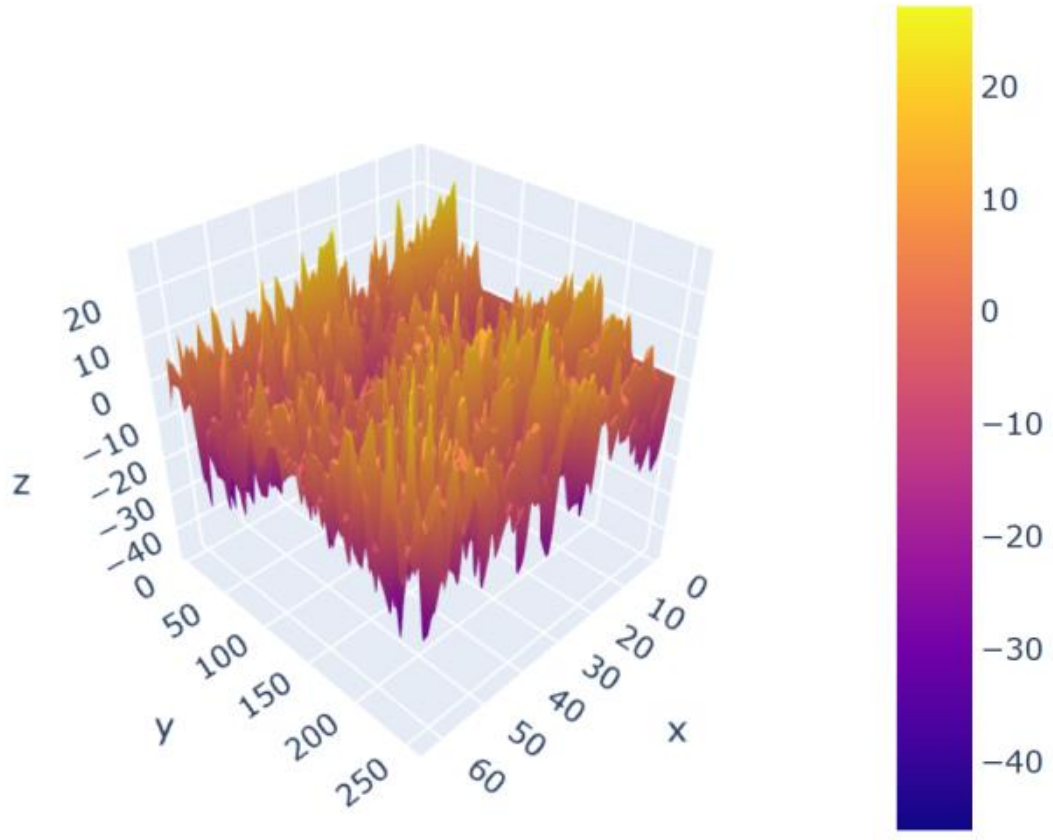

d

Fig. 2. Dataset visualization. X-axis for the number of electrode (from 1 to 64), Y-axis - the record of value for each electrode (256 times with a length of 1 second), Z-axis - the signal amplitude, µV, a -1.alk_s1, b - 2.1alk_s2, c - 3.not_alk_s1, d - 4.not_alk_s2

The ratio of the electrodes on the head and the X-axis (fig.2) is presented in table 2.

Table 2. Location of the position of the electrodes on the X-axis for Fig. 1, Fig. 2

| Electrode number in dataset | 1 | 2 | 3 | 4 | 5 | 6 | 7 | 8 | 9 | 10 | 11 | 12 | 13 | 14 | 15 | 16 |
|---|---|---|---|---|---|---|---|---|---|---|---|---|---|---|---|---|
| The name of the electrode in Fig. 1 | FP1 | FP2 | F7 | F8 | AF1 | AF2 | FZ1 | F4 | F3 | FC6 | FC5 | FC2 | FC1 | T8 | T7 | CZ |

| Electrode number in dataset | 17 | 18 | 19 | 20 | 21 | 22 | 23 | 24 | 25 | 26 | 27 | 28 | 29 | 30 | 31 | 32 |
|---|---|---|---|---|---|---|---|---|---|---|---|---|---|---|---|---|
| The name of the electrode in Fig. 1 | C3 | C4 | CP5 | CP6 | CP1 | CP2 | P3 | P4 | PZ | P8 | P7 | PO2 | PO1 | O2 | O1 | X |

| Electrode number in dataset | 33 | 34 | 35 | 36 | 37 | 38 | 39 | 40 | 41 | 42 | 43 | 44 | 45 | 46 | 47 | 48 |
|---|---|---|---|---|---|---|---|---|---|---|---|---|---|---|---|---|
| The name of the electrode in Fig. 1 | AF7 | AF8 | F5 | F6 | FT7 | FT8 | FPZ | FC4 | FC3 | C6 | C5 | F2 | F1 | TP8 | TP7 | AFZ |

| Electrode number in dataset | 49 | 50 | 51 | 52 | 53 | 54 | 55 | 56 | 57 | 58 | 59 | 60 | 61 | 62 | 63 | 64 |
|---|---|---|---|---|---|---|---|---|---|---|---|---|---|---|---|---|
| The name of the electrode in Fig. 1 | CP3 | CP4 | P5 | P6 | C1 | C2 | PO7 | PO8 | FCZ | POZ | OZ | P2 | P1 | CPZ | nd | Y |

This dataset has some artifacts. In fig. 2 - c there is a sudden increase in tension, which can be caused by eye movement or blinking. To exclude this kind of artifact, the principal component method was used. The method of principal components is a multidimensional statistical analysis method used to reduce the dimension of the feature space with minimal loss of useful information. The potential of electrooculography (EOG) is one of the most popular artifacts that occur with eye movement. In this case, the maximum amplitude of artifacts is observed in the frontal leads and decreases towards the occipital leads. In the next researches for [18,19,20] to remove artifacts caused by involuntary eye movements of the subject from a multi-channel EEG, a wide analysis of the main components is used. For these artifacts, it is very difficult to visually find regularity in the presented figures (Fig. 2). Therefore, it is advisable to try neural networks and machine learning for EEG signal recognition.

### 3.2.1. Wavelet transforms

We considered the most popular methods of signal preprocessing - wavelet transform and decomposition into a fast Fourier series.

In many researches, the fast Fourier transform is used in conjunction with the wavelet transform [25, 26, 27, 28, 29].

The wavelet transform carries a huge amount of information about the signal, but, on the other hand, has a strong redundancy, since each point of the phase plane affects its result.

A continuous wavelet transform is defined as the scalar product of the original signal x(t) and the daughter wavelet function $¥_{t,a}(t)$:

$$W(\tau, a) = (x(t), ¥_{\tau,a}(t)) = \int_{-\infty}^{+\infty} x(t)¥t^{*}_{\tau,a}(t)dt, \tau \in R^{+}, \quad (1)$$

where W(τ,a) - wavelet expansion coefficients; τ, a – пparameters of time shift and scale, respectively; operator * means complex pairing.

Child wavelet functions $¥_{\tau,a}$ , formed by shear and scale operations of the mother wavelet function $¥_{\tau}$ and related to it by the ratio:

$$¥_{\tau,a}(t) = \frac{1}{\sqrt{a}}¥(\frac{t-\tau}{a}) \quad (2)$$

The complex Morlet wavelet, which is the product of a complex sinusoid and a Gaussian, is used as the mother wavelet function. The analytical expression of the Morlet wavelet has the form:

$$¥(t) = \frac{1}{\sqrt[4]{\sigma^2 * \pi}} * e^{-\frac{t^2}{2\sigma^2}} * e^{jw_0 t}, \quad (3)$$

where ω0 – maternal wavelet center frequency; σ – standard deviation of the envelope of the mother wavelet.

Using a stationary wavelet transform, the signal from the electrode to 4 frequency levels delta (0-4 Hz), theta (4-8 Hz), alpha (8-12 Hz), beta (12-20 Hz) were decomposed, fig.3.

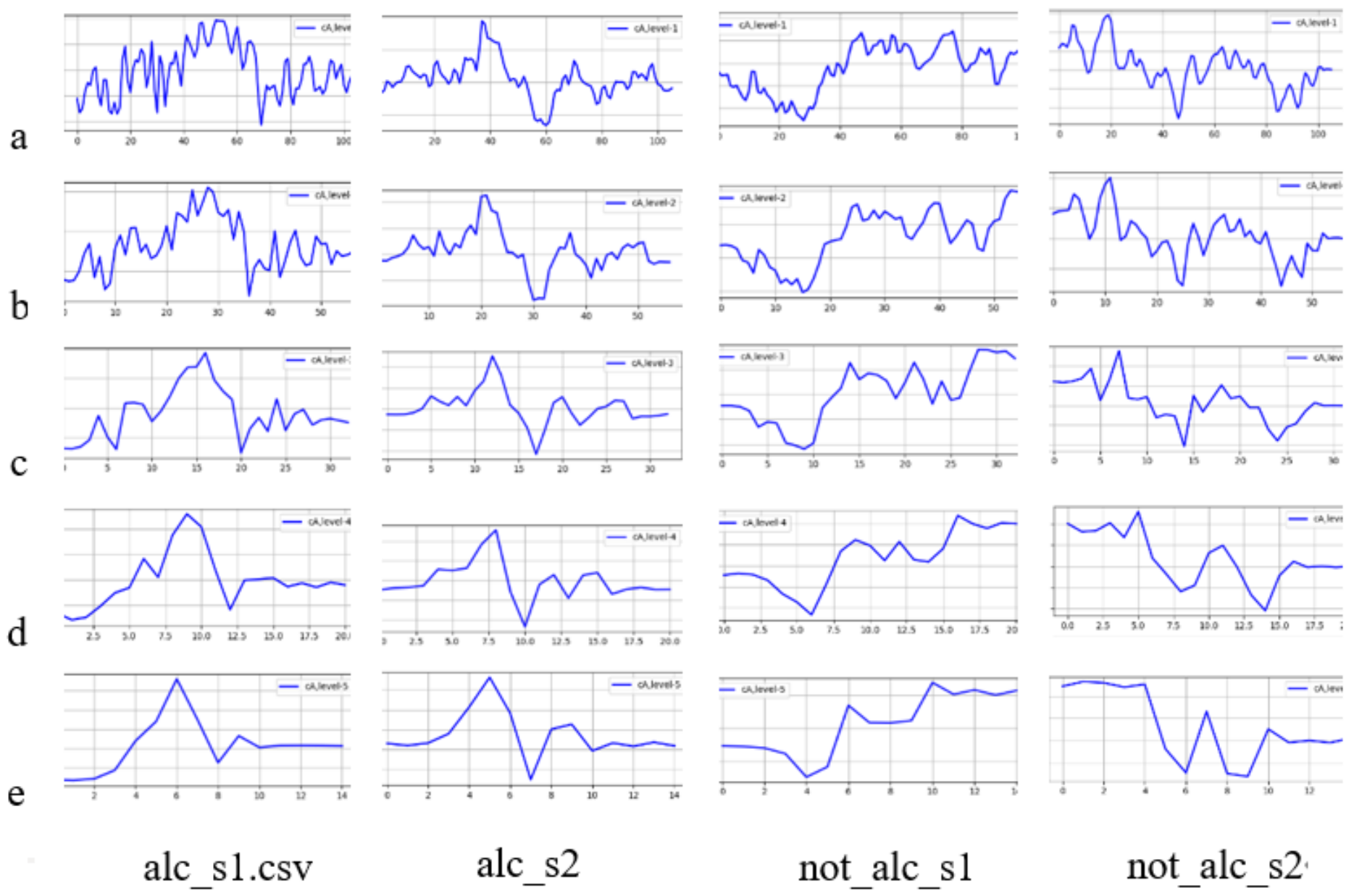

Fig. 3. a - initial signal, b - beta (12-20 Hz), c - alpha (8-12 Hz), d - theta (4-8 Hz), e - delta (0-4 Hz),

The received image dataset will be used as input data to the neural network.

### 3.2.2. Fast Fourier Series

The continuous Fourier transform, and the discrete Fourier transform have not found wide application in the process of extracting attributes due to their low efficiency, which was explained in the next articles [30,31,32,33,34].

The most popular is the decomposition of the signal into harmonic components using the Fast Fourier transform.

For the signal x (n), presented in the form of a sequence of samples, taken with sampling frequency Fs, time moments with numbers n = 0,1, ..., N-1, the discrete Fourier transform is defined as:

$$F(k) = \sum_{n=0}^{N-1} x(n) * e^{-\frac{2\pi j}{N}kn}, k = 0,1,\dots,N-1, \quad (4)$$

where F (k) is the complex amplitude of the sinusoidal signal with a frequency k * $\Delta$ f, $\Delta$ f = Fs / N resolution (step) in frequency, x (n) are the measured signal values at time instants with numbers n = 0,1, ... .N -1.

The result of expanding the signal into a Fast Fourier series with the scipy.fftpack library is shown in fig.4.

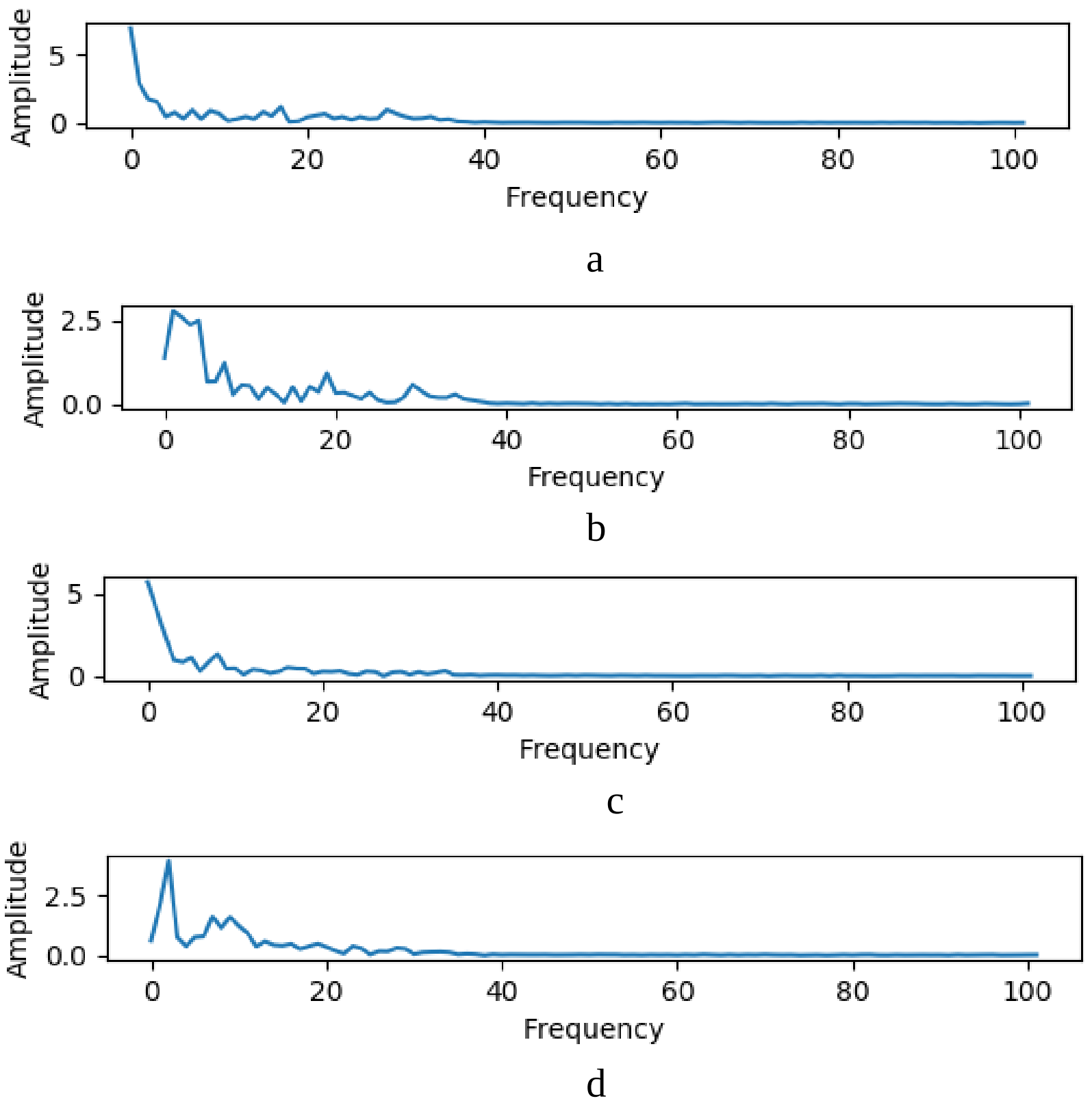

Fig.4. Expanding the signal into a Fast Fourier series. a -1.alk_s1, b -2.1alk_s2, c - 3.not_alk_s1, d - 4.not_alk_s2

The received image dataset will be used as input data to the neural network.

## 3. Experimental research

### 3.1 Machine Learning

The use of machine learning in classification tasks today is becoming less popular due to the development of deep neural networks. But in the next papers, results with high accuracy with machine learning were obtained [21,22,23,24]. For this reason, we tried using machine learning. In our research for machine learning tasks, the available dataset of 480 excel files was converted to a single file, of the following form, table. 3

Table. 3. Type of machine learning dataset

| | sensor position | sensor value | ... | channel | subject identifier |
|---|---|---|---|---|---|
| 0 | 0 | -3.550 | ... | 0 | 1 |
| 1 | 0 | -5.015 | ... | 0 | 1 |
| 2 | 0 | -5.503 | ... | 0 | 1 |
| ... | ... | ... | ... | ... | ... |
| 7831549 | 63 | -9.054 | ... | 63 | 0 |
| 7831550 | 63 | -9.054 | ... | 63 | 0 |
| 7831551 | 63 | -9.054 | ... | 63 | 0 |

[7831552 rows x 5 columns]

Algorithms Logical Regression, Naive bayes, k-Nearest Neighbors, Support_Vector_Machines shows results of approximately 0.50 accuracy. Algorithms Random Forest Classifier - 0, 75 accuracies. The maximum result was obtained with the use of Classification and Regression Trees (CART) - 0.81 accuracies.

### 3.2. Deep neural networks

To increase accuracy, we decided to use deep neural networks. We decided to use a convolutional neural network (CNN) that works with images.

Figure 5 shows the two-dimensional (2D) graphics data correlation (Python3.7, matplotlib).

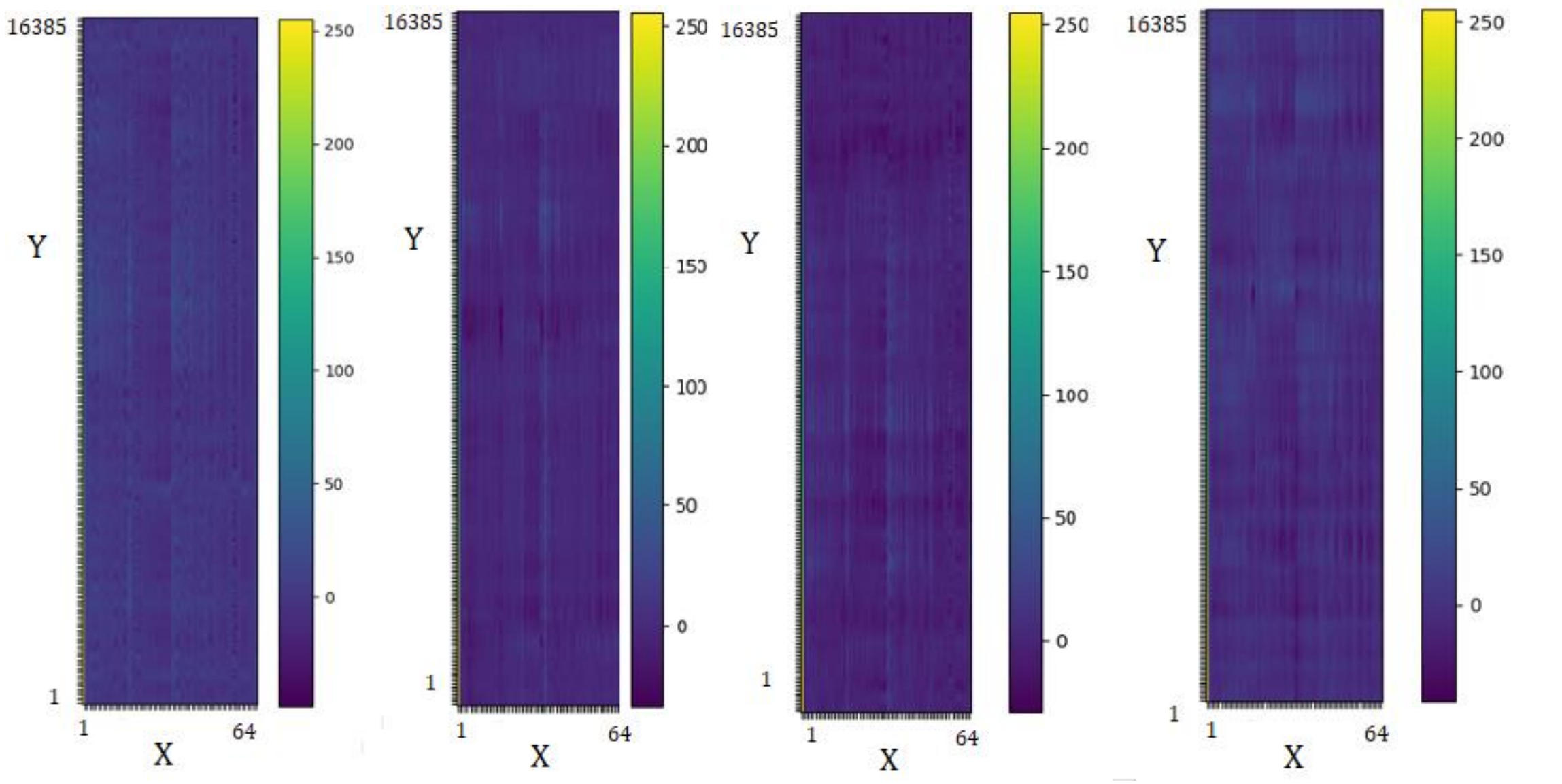

***a*** ***b*** ***c*** ***d***

Fig.5. Correlation EEG signals. Where, X-axis for the electrode from 1 to 64, Y-axis – value for electrode. Every electrode has 256 records with length 1 sec. a -1.alk_s1, b -2.1alk_s2, c - 3.not_alk_s1, d - 4.not_alk_s2

Image analysis shows a high correlation between regions that are close to each other. Visually we noticed that brain regions show different correlation values between subjects for the following regions PO3-CPZ and F4-C4. Visual observations allow us to conclude that the image data can be used for deep machine learning in the classification task.

Today, the following CNN can be used in the EEG signal classification process [35,36,37,38], fig.6.

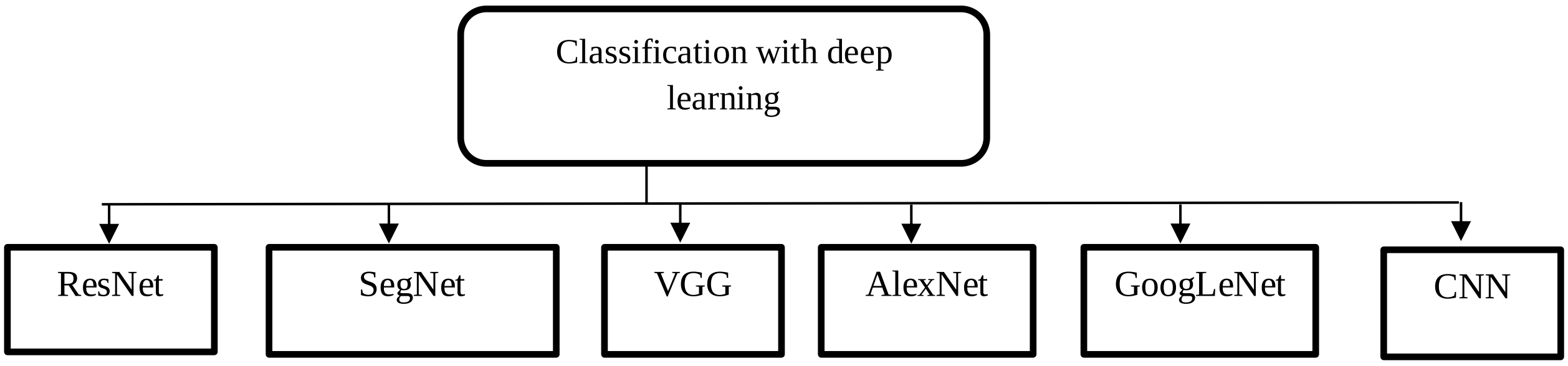

Fig.6. Popular deep neural networks for image classification

Initially, for classification tasks, we used the Inception network. The weights originally trained for layers 1 through 205 were used. Further, the network was completed on training images. The result of the classification accuracy was 0.72. The low accuracy is since the scales were initially trained to classify full-fledged images and not graph. Similar results were obtained for networks - Resnet, VGG.

Therefore, for classification, the new CNN was created. CNN consists of 6 layers of Convolution_2D and MaxPooling2D layers after the second and fourth convolution. On all layers except the output fully connected layer, the ReLU activation function is used, the last layer uses softmax. To regularize our model, after each subsample layer and the first fully connected layer, the Dropout layer was used.

The following graphs were used as input images:

- source graphics without conversion;
- correlation graphs;
- Beta charts;
- Alpha charts;
- Theta graphics;
- Delta charts;
- Fast Fourier Series.

Correlation graphs showed the highest accuracy.

A schematic network when working with correlation graphs is shown in fig. 7.

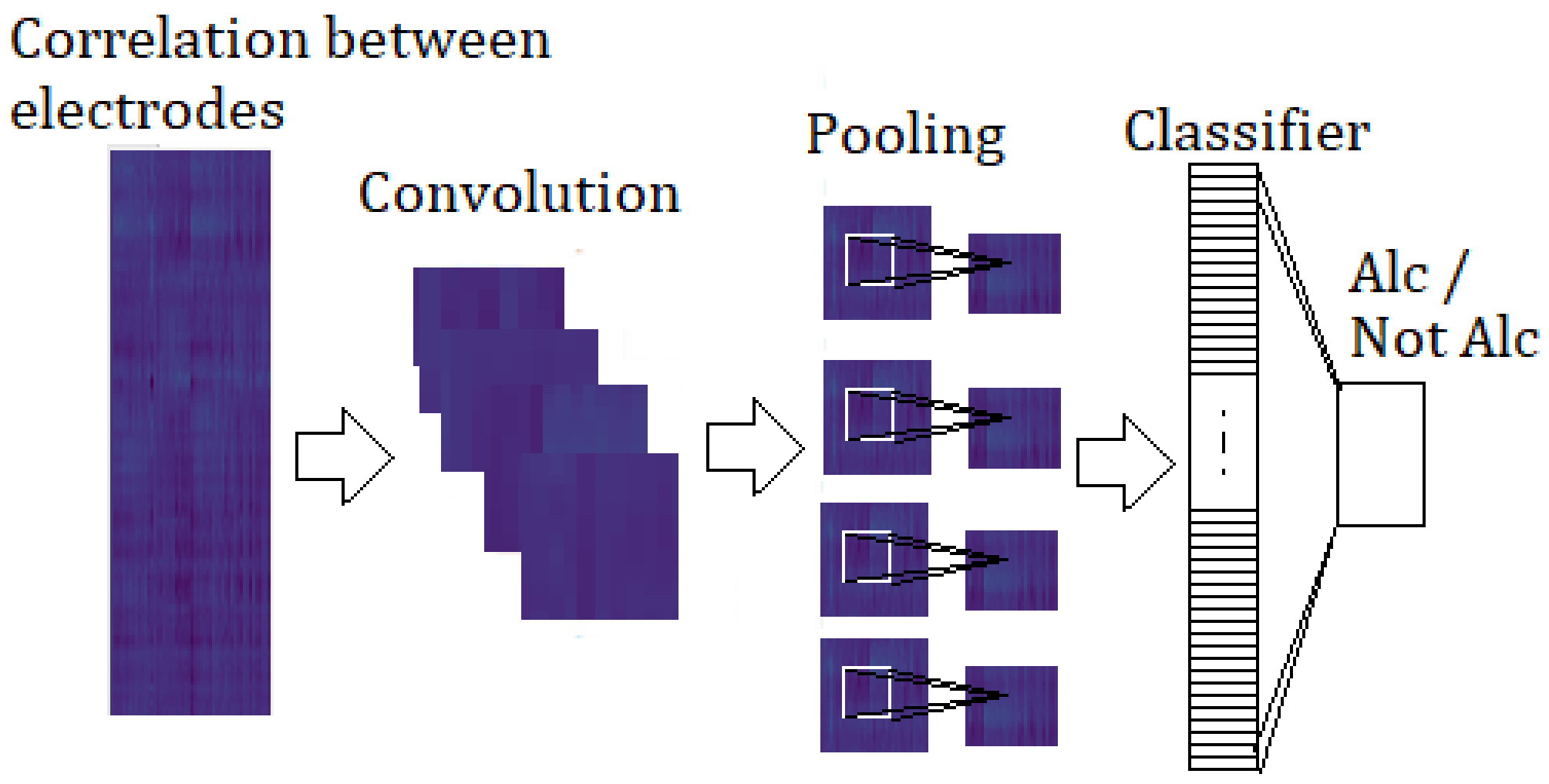

Fig.7. Schematic representation of a neural network when working with correlation graphs

The results of the classification accuracy in % when using various input images are presented in table 4.

Table 4. Accuracy of CNN model for different dataset

| Developed CNN | Beta | Alpha | Theta | Delta | Correlation EEG | Fast Fourier Series. | Source graphics without conversion |
|---|---|---|---|---|---|---|---|
| Classification accuracy,% | 85 | 82 | 75 | 56 | 92 | 75 | 86 |

For the developed CNN model, the higher accuracy for classification objects between alcohol and not alcohol objects when we use Correlation EEG images was received.

## 4. Discussion and conclusions

The highest accuracy result in classification was obtained using an image with the correlation of signals. The next areas PO3-CPZ and F4-C4 have the highest correlations.

From the frequency range, a high classification result when working with the Beta range was obtained. The delta range showed the lowest result, which is associated with the loss of the useful signal in the original frequency. Using images with the Fourier series has accuracy commensurate with accuracy for the machine learning algorithm.

In this research, there is no direct pattern between the magnitude of the voltage across the electrode and the group of studies. In 72% of cases, it is observed that in the group of alcoholics the voltage on the electrode is lower than in the control group. But, the location of the electrodes is different and there is no way to establish an exact relationship between the magnitude of the voltage on the electrode and the presence of the disease.

In many papers, the beta, alpha, and theta rhythms of the EEG signal were used for classification signals. In this research, it is shown that with the same type of data about the object under research (only electrode voltage), it is preferable to use convolution networks with images of the correlation of EEG signals.

For correct research in the field of analysis of EEG data using neural networks, it is necessary to submit as much data as possible from studies: age, gender, medical history, etc.

Much research in the field of alcohol recognition by EEG signals has different results. In order to avoid it, it is necessary to develop a standard in the field of using neural networks regulating the number of signs for neural networks for classifying an alcoholism disease in a subject.

**Dataset and code:** https://github.com/Ildaron/3.eeg_recognation

**Conflicts of Interest**: None
**Funding**: None
**Ethical Approval**: Not required